\documentclass[12pt]{article}
\def\cF{{\cal F}}

\def\cH{\cal H}

\newfont{\goth}{eufm10 scaled \magstep1}

\def\a{\alpha}

\def\b{\beta}

\def\c{\gamma}
\def\d{\delta}

\def\r{\rho}
\def\s{\sigma}\def\S{\Sigma}
\def\t{\tau}
\def\th{\theta}

\def\beq{\begin{equation}}\def\eeq{\end{equation}}
\def\beqa{\begin{eqnarray}}\def\eeqa{\end{eqnarray}}
\def\barr{\begin{array}}\def\earr{\end{array}}

\def\del{\partial}

\def \ys {{y\kern-.5em / \kern.3em}}

\def\nn{\nonumber}
\def\bd{\begin{document}}
\def\ed{\end{document}}
\def\ba{\begin{array}}
\def\ea{\end{array}}
\def\bea{\begin{eqnarray}}
\def\eea{\end{eqnarray}}
\def\ft#1#2{{\textstyle{{\scriptstyle #1}\over {\scriptstyle #2}}}}
\def\fft#1#2{{#1 \over #2}}
\newcommand{\be}{\begin{equation}}
\newcommand{\ee}{\end{equation}}
\newcommand{\eq}[1]{(\ref{#1})}
\def\eqs#1#2{(\ref{#1}-\ref{#2})}
\def\det{{\rm det\,}}
\def\tr{{\rm tr}}
\newcommand{\ho}[1]{$\, ^{#1}$}
\newcommand{\hoch}[1]{$\, ^{#1}$}
\def\ra{\rightarrow}
\def\uha{{\hat {\underline{\a}} }}
\def\uhc{{\hat {\underline{\c}} }}

\def \Om {\Omega}
\def \bfd {{\bf d}}
\def \del {\partial}
\def \eps {\epsilon}
\def \Z {{\bf Z}}
\def \xb {\bar{x}}
\def \la {\langle}
\def \ra {\rangle}
\def \Omt {\tilde \Omega} 
\def \la {\langle}
\def \ra {\rangle}

\begin{document}

\hfill{NEIP-98-022}

\hfill{hep-th/9812219}

\vspace{20pt}

\begin{center}

{\Large\bf Noncommutative Open String and D-brane}
\vspace{30pt}

{\large Chong-Sun Chu\hoch{1} and Pei-Ming Ho\hoch{2}}

\vspace{15pt}

\begin{itemize}
\item[$^1$] {\small \em Institute of Physics,
University of Neuch\^atel, CH-2000 Neuch\^atel, Switzerland} 
\item[$^2$] {\small \em Department of Physics, National Taiwan
University, Taipei 10764, Taiwan, R.O.C.}
\end{itemize}

\vskip .2in
\sffamily{cschu@sissa.it \\
pmho@phys.ntu.edu.tw}

\vspace{60pt}

{\bf Abstract}
\end{center}

In this paper we consider the quantization of open strings
ending on D-branes with a background $B$ field.
We find that spacetime coordinates of the open string end-points
become noncommutative, and correspondingly the D-brane worldvolume
also becomes noncommutative. This provides a string theory
derivation and generalization of the noncommutativity obtained
previously in the M(atrix) model compactification.
For D$p$-branes with $p \geq 2$ our results are new
and agree with that of  M(atrix) theory for the case of $A=0$ (where $A$ is
the worldvolume gauge field)
if the T-duality radii are used. 

\newpage

\section{Introduction}

M(atrix) theory \cite{BFSS} compactified on a torus is described by a
supersymmetric Yang-Mills (SYM) theory living on the dual torus
\cite{BFSS,taylor}. However,
this simple picture no longer holds when a background field is present.
It was shown by Connes, Douglas and Schwarz in \cite{CDS} that
for the M(atrix) model compactified on a $T^2$, 
noncommutative SYM arises naturally if
there is a background field $C_{-12}$. This is justified
from the D-string point of view \cite{DH} by dualizing one
cycle of the torus.. Later, it was demonstrated directly how
noncommutativity arises from the  D0-brane point of view \cite{CK,KO}.

It was then also suggested in \cite{HV2} by Hofman and Verlinde
that a D-brane worldvolume is noncommutative already in
string theory before taking the M(atrix) model limit.
Taking a similar approach as in \cite{AAS},
we demonstrate in this paper that the noncommutativity on
D-brane worldvolume has an explanation in terms of open string
quantization in background fields.

We find that the end-points of the string have
noncommutative coordinates and therefore we infer that
the D-brane worldvolume is a noncommutative space.
The noncommutativity derived from string quantization
is shown to agree with the results of M(atrix) model
compactified on torus when the $U(1)$ field strength
$F$ on the D-brane worldvolume vanishes.
In general the noncommutativity is determined by
the gauge invariant combination $\cF=(B-F)$, instead of
just the NS two-form $B$ field.
We show that the D-brane worldvolume does not
need to be a compact space in order to have noncommutativity.

We also give explicit formula telling  
how the string theory data (the gauge invariant background 
field strength $\cF$) should be encoded in the noncommutative
worldvolume of a D$p$-brane for $p \geq 2$, generalizing previous results.

\section{Classical Action}

Consider a fundamental string ending on a D$p$-brane,
the bosonic part of the action is \cite{DLP,Lei}
\be \label{action}
S_B= {1 \over 4\pi\alpha'} \int_{\Sigma} d^2\sigma
\bigl[ g^{\a\b}G_{\mu\nu} \partial_\a X^{\mu}\partial_\b X^{\nu}+
\eps^{\a\b} B_{\mu\nu}\partial_\a X^{\mu}\partial_\b X^{\nu} ]
+ {1 \over 2\pi\alpha'}\oint_{\partial \Sigma} d \tau A_i(X)
\partial_{\tau}X^i, 
\ee
where $A_i,\ i=0,1,\cdots, p$, is the $U(1)$ gauge field
living on the D$p$-brane.
Here the string background is 
\be
G_{\mu\nu} = \eta_{\mu\nu}, \quad \Phi =\mbox{constant}, 
\quad H=dB=0.
\ee
 
Variation of the action yields the equations of motion for a
free field 
\be
(\del^2_{\tau}-\del^2_{\s}) X^\mu =0
\ee
and the following boundary conditions at $\s =0, \pi$:
\bea
&\del_\s X^i + \del_\tau X^j \cF_j{}^i =0, \quad i,j= 0,1,\cdots, p,
\label{BC1}\\
&X^a =x_0^a,\quad a =p+1, \cdots, 9. \label{BC2}
\eea
Here
\be
\cF =B-dA=B-F
\ee 
is the modified Born-Infeld field strength
and $x_0^a$ is the location of the D-brane.
Indices are raised and lowered by $\eta_{ij} = (-, +, \cdots, +)$.

If both ends of a string are attached to the same D$p$-brane,
the last term in (\ref{action}) can be written as
\be
\frac{-1}{4\pi\a'}\int_{\Sigma}d^2\s
\eps^{\a\b}F_{ij}\del_{\a}X^i\del_{\b}X^j.
\ee
Furthermore consider the case $B=\sum_{i,j=0}^{p}B_{ij}dX^i dX^j$,
then the action (\ref{action}) can be written as
\be
S_B=-\int d\t L=\frac{1}{4\pi\a'}\int d^2\s
\bigl[ g^{\a\b}\eta_{\mu\nu}\partial_\a X^{\mu}\partial_\b X^{\nu}+
\eps^{\a\b}\cF_{ij}\partial_\a X^{i}\partial_\b X^{j} ].
\ee
Note that for open strings $B$ and $F$ always
appear together in the combination $\cF=B-F$,
which is invariant under both gauge transformations for
the one-form gauge field $A$
\be
A\rightarrow A+d\Lambda, \quad B\rightarrow B,
\ee
and for the two-form gauge field $B$
\be
B\rightarrow B+d\Lambda, \quad A\rightarrow A+\Lambda.
\ee

We will be interested in the consequence of \eq{BC1} due to the
presence of constant background field $\cF$ along
the directions of the worldvolume of the D-brane.
A remark about the boundary
conditions is in order. In the usual boundary state formalism, a
D-brane is described by the boundary state \cite{VFPSLR,AS1}.
The boundary conditions
for the open string ending on the D-brane are translated into
conditions (the overlapping conditions) satisfied  by the boundary
state. In particular, the boundary conditions \eq{BC1} and
\eq{BC2} are not implemented as operator constraints,  but rather
implemented as constraints on the space of boundary states.

In this paper, we are interested in seeing the origin of 
the noncommutativity on
the D-brane worldvolume field theory due to the presence of
non-trivial two-form background field $\cF$. Instead of using the boundary
state formalism,  we will use a more direct operator approach by
implementing the conditions \eq{BC1} and \eq{BC2} as operator constraints
and investigate their consequences. This approach is
intuitively clearer and easier to make connection with previous results
\cite{CDS,DH,HW3,Ho} on noncommutative gauge theory
as most results in the literatures are expressed in the
operator language. 
Modulo technical details, we expect that the two
approaches give completely equivalent results because of the
standard duality between operators and states.  It would be very
interesting to see explicitly how noncommutativity arises in the
boundary state formalism.

The general solution of $X^k$ to the equations of motion is 
\be \label{mode}
X^{k}=x_0^k+(a_0^k \tau + b_0^k \sigma)
+ \sum_{n\neq 0} {e^{-in\tau} \over n}
\bigl(ia^k_n \cos n\sigma + b_n^k \sin n\sigma \bigr).
\ee
Substituting \eq{mode} into the boundary conditions \eq{BC1},
we get
\be
b_n^k +  a_n^j \cF_j{}^k =0, \quad  \mbox{for all $n$}.
\ee
Eliminating $b_n^k$ and denoting $a_0^k = p_0^k$, we get
\be \label{mode1}
X^k =x_0^k+(p_0^k \tau -  p_0^j \cF_j{}^k \sigma)+ 
\sum_{n\neq 0} {e^{-in\tau} \over n}
\bigl(ia^k_n \cos n\sigma -   a_n^j \cF_j{}^k \sin n\sigma \bigr).
\ee
Similarly we can solve the equations of motion and boundary conditions
for $X^a$ and get
\be \label{mode2}
X^a=x_0^a+b^a\s+\sum_{n\neq 0}\frac{e^{-in\t}}{n}a^a_n \sin n\s,
\quad a=p+1,\cdots, 9,
\ee
where $x_0^a+\pi b^a$ is the location of the D-brane to which
the other end-point of the open string is attached.
If both ends of the string end on the same brane then $b^a=0$.

The canonical momentum from the action \eq{action} is given by
\bea
2\pi\a' P^k(\t,\s)&=&\del_\tau X^k + \del_\sigma X^j\cF_j{}^k,
\label{mom} \\
2\pi\a' P^a(\t,\s)&=&\del_\tau X^a.
\label{mom'}
\eea
Substituting \eq{mode1} into the above expressions, we get
\bea
2\pi\a' P^k(\t,\s)&=&\{ p_0^l +
\sum_{n\neq 0} a_n^l e^{-in\tau} \cos n\sigma \}M_l{}^k,
\label{mode3} \\
2\pi\a' P^a(\t,\s)&=&-i\sum_{n\neq 0}e^{-in\t}a^a_n\sin n\s,
\label{mode4}
\eea
where
\be \label{M}
M_{ij}=\eta_{ij}-\cF_i{}^k\cF_{kj}.
\ee

One can check that the total momenta
\be
P_{tot}^k(\t) = \int_0^\pi d\s P^k(\t,\s)=\frac{1}{2\a'}p_0^l M_l{}^k,
\ee
are constants of motion.

The center of mass coordinates of the string is
\be
x_{cm}^k(\t)=\frac{1}{\pi}\int_0^{\pi}d\s X^k(\t,\s).
\ee
It is
\be
x_{cm}^k=x_0^k+p_0^k\t-\frac{\pi}{2}p_0^j\cF_j{}^k
-\sum_{n\neq 0}\frac{e^{-in\t}}{n^2}(1-(-1)^n)a^j_n\cF_j{}^k.
\ee

The Hamiltonian is defined by
\be
H_B=\int d\s P_{\mu}X^{\mu} - L
\ee
and is found to be
\be
H_B=\frac{1}{4\pi\a'}\int d\s
\left( (\del_{\t}X)^2+(\del_{\s}X)^2\right).
\ee
Using the mode expansion (\ref{mode1}),(\ref{mode2}),
we find
\be \label{Ham}
H_B=\frac{1}{4\a'}\left(M_{ij}p_0^i p_0^j+b^a b^a+\sum_{n\neq 0}
(M_{ij}a^i_n a^j_{-n}+a^a_n a^a_{-n})\right).
\ee
 
\section{Quantization} \label{quant}

The quantization of $X(\t,\s)$ has to be different from
the usual canonical commutation relations for free fields
because the standard equal time commutation relations are  
inconsistent with the boundary conditions (\ref{BC1})
\cite{SJ,AAS}.
To see this, we use (\ref{BC1}) and (\ref{mom}) and obtain
\be
2\pi\a' P^k(\t,0)\cF_k{}^i=-\del_{\s}X^j(\t,0)M_j{}^i.
\ee
It follows that
\be \label{PP}
2\pi\a'[P^k(\t,0),P^j(\t,\s)]\cF_k{}^i=
-[\del_{\s}X^k(\t,0),P^j(\t,\s)]M_k{}^i.
\ee
This relation makes it impossible to impose both
the following standard canonical commutation relations 
consistently,
\bea
&[X^i(\t,\s),P_j(\t,\s')]=i\d^i_j\d(\s-\s'), \\
&[P_i(\t,\s),P_j(\t,\s')]=0.
\eea
A similar situation also occurred in the usual quantization of Maxwell
field in the Coulomb gauge. There one finds that the gauge fixing
condition is not consistent with the standard canonical quantization
and one has to modify the quantization in a consistent manner. We will 
do the same in the following. 
We expect that our quantization procedure to be 
equivalent to the constraint quantization of Dirac.
\footnote{We are grateful to M. Cederwall for informing us that 
this is indeed the case.}
The conflict between the usual relations for free fields
in this case was first realized in \cite{SJ,AAS}.
However their quantization does not agree with ours. (See
in particular eqn. \eq{AASJ} below and the discussions before it.).
 
Let us first review the usual quantization procedure and then we will
see that there is a natural generalization of the usual canonical
quantization procedure that takes care of our case. 
The usual way to quantize a classical system is to start with 
the symplectic structure on the phase space. For example, in the
standard string case with $B=A=0$, one first rewrites the 
action in the Hamiltonian form
\be
-\int d^2\s(P_{\mu}\dot{X}^{\mu}-\cH),
\ee
where $\cH$ is the Hamiltonian density.
The symplectic form
\be\label{omega}
\Omega(\mbox{fields})= \int d\s  \bfd P_{\mu} \bfd{X}^{\mu}
\ee
is extracted from the first piece \cite{Witten} and it defines
the standard Poisson bracket for the fields $X^\mu$, $P^\mu$
\bea
&[X^\mu(\t,\s),P_\nu(\t,\s')]_{PB}=\d^\mu_\nu \d(\s-\s'), 
\label{poisson1}\\
&[P_\mu(\t,\s),P_\nu(\t,\s')]_{PB}=0, \quad 
[X_\mu(\t,\s),X_\nu(\t,\s')]_{PB}=0, \label{poisson2}
\eea
which are the standard equal time commutation relations
for free fields.
Plugging in the mode expansions \eq{mode1}, \eq{mode2}, \eq{mode3} and
\eq{mode4} of $X^\mu$ and $P^\mu$ for the standard case of $B= A =0$,
one gets the usual commutation relations for the modes. 

One can also derive the Poisson structure for the modes directly
without going through the fields. To do this, one just have to 
follow the above steps by
first plugging in the mode expansions for $X^\mu$ and $P^\mu$ and 
then evaluate \eq{omega}. The result is  
\be \label{modeomega}
\Om(\mbox{modes}) =\frac{1}{2\a'}\left\{\eta_{ij}\bfd p_0^i
\bfd x_0^j +
\sum_{n > 0}\frac{-i}{n}(\eta_{ij}\bfd a^i_n\bfd a^j_{-n}
+\bfd a^a_n\bfd a^a_{-n})\right\}.
\ee
This implies the same standard commutation relations for the  modes and
the two approaches are completely equivalent. 

All of these are quite standard. In our case, because of the
boundary conditions \eq{BC1},
we saw that it is inconsistent to impose \eq{poisson1}, \eq{poisson2}, 
even in the Poisson limit. 
We propose to  use
\be \label{omegat}
\Omt(\mbox{modes}) = \la\int d\s\bfd P_{\mu}\bfd X^{\mu} \ra, 
\ee
where $\la\cdot\ra$ is the time average
\be
\la O \ra = \lim_{T\rightarrow \infty} \frac{1}{2T} \int_{-T}^T O d\tau 
\ee
as the definition of the symplectic structure for  the modes in general.
A remark on this definition is in
order.  Usually one takes the symplectic form to be given by $\int d\s
\bfd P_{\mu}\bfd X^{\mu}$ because the mode expansion of $X$ is given
by an orthogonal basis of spatial functions.  However, generically the
modes are orthogonal only as functions on the whole spacetime, but not
necessarily orthogonal in spatial dependence.  In our case, due to the
boundary conditions, the spatial dependence of the mode expansion is
not orthogonal, so if we do not average over time we will be left with
$\tau$-dependence in the symplectic form for the modes. This is not
consistent.  Here we talk about string as an example, but in general
one can use \eq{omegat} for a generic system. It is easy to see
that \eq{omegat} reproduces the usual results for all
typical cases in quantum field theory. 
Applying to our particular case with the string defined by
\eq{action}-\eq{BC2}, eq.\eq{omegat} 
defines a consistent quantization.
{}From another viewpoint, 
the definition (\ref{omegat}) is very natural
because the time integration is already there in the action
from which the symplectic form is extracted.
We will give further justifications of our 
results later by showing that in our quantization,
the correct time evolution (\ref{HX})   of $X$ 
and the standard Virasoro algebra is obtained.

Now substituting (\ref{mode1}) and (\ref{mom}), we find
\be \label{modeF}
\Omt=\frac{1}{2\a'}\left\{M_{ij}\bfd p_0^i
(\bfd x_0^j+\frac{\pi}{2}\cF^j{}_k\bfd p_0^k)+
\sum_{n > 0}\frac{-i}{n}(M_{ij}\bfd a^i_n\bfd a^i_{-n}
+\bfd a^a_n\bfd a^a_{-n})\right\}.
\ee
As a consistency check, it is easy to see that
this reduces to \eq{modeomega} in the usual situation of
$\cF=0$.
This expression can be simplified if we shift the range
of $\s$ by $\pi/2$ so that $\s\in[-\pi/2,\pi/2]$ and
define the coordinates for the center of a string
\be
\xb_0^{i}=x_0^i+\frac{\pi}{2}\cF_{ij}p_0^j.
\ee
It implies the following commutation relations
\bea
&[\xb_0^{i}, \xb_0^{j}]=[p_0^i, p_0^j]=0, \\
&[\xb_0^{i}, p_0^j]=i2\a'M^{-1ij} ,\label{cr1} \\
&[a_n^i, \xb_0^{j}]=[a_n^i, p_0^j] =0, \\
&[a_m^i, a_n^j]=2\a'mM^{-1ij}\d_{m+n}, \label{cr2} \\
&[a_m^a, a_n^b]=2\a'm\d^{ab}\d_{m+n}.
\eea
In terms of $x_0^i$ it is
\bea
&[x_0^i, p_0^j]=i2\a'M^{-1ij}, \\
&[x_0^i,x_0^j]= i 2\pi\a'(M^{-1}\cF)^{ij},\\
&[p_0^i, p_0^j]=0.
\eea
Reality of the fields $X^\mu$ implies that 
$x_0^\mu, b^a, p_0^k$ are real and
\be \label{inv}
a_n^\mu{}^\dagger = a_{-n}^\mu.
\ee
As a consistency check, it is easy to see that the above commutation
relations respect this reality structure.

Using (\ref{Ham}) we can check that
\be \label{HX}
[H,X^\mu(\t,\s)]=-i\del_{\t}X^\mu(\t,\s), \quad 
[H,P^\mu(\t,\s)]=-i\del_{\t}P^\mu(\t,\s),
\ee
and that the center of mass coordinate $x_{cm}^k$ is conjugate
to the total momentum $P^k_{tot}$ in the usual sense
\bea
&[x_{cm}^j,P^k_{tot}]=i \eta^{jk}, \\
&[x_{cm}^j,x_{cm}^k]=0, \quad [P^j_{tot},P^k_{tot}]=0.
\eea

The spacetime coordinate $X(\t,\s)$ and momentum $P(\t,\s)$
are derived concept and their equal time commutators can be derived.
It is 
\bea
&[P^i(\t,\s),P^j(\t,\s')]=0,  \\
&[X^k(\t,\s),X^l(\t,\s')]=-2i\a'(M^{-1}\cF)^{kl}\left[
(\s+\s'-\pi)+\sum_{n\neq 0}\frac{1}{n}\sin n(\s+\s')\right],  
\label{XX} 
\eea
\be
 [X^i(\t,\s),P^j(\t,\s')]= i \eta^{ij}\cdot \frac{1}{\pi}
\left(1+ \sum_{n\neq 0}\cos n\s \cos n\s' \right). \label{XP}
\ee

Letting $\th=(\s+\s')$,
the infinite series on the right hand side of (\ref{XX})
is the Fourier expansion
\be
\sum_{n\neq 0}\frac{1}{n}\sin n\th=\th-\pi
\ee
for $\th\in(0,2\pi)$.
But at the boundary we have
\be
\sum_{n\neq 0}\frac{1}{n}\sin n\th=0
\ee
for $\th=0, 2\pi$.
Therefore
\be
[X^k(\t,\s),X^l(\t,\s')]=0
\ee
for all values of $\s$ and $\s'$ except that when
$\s=\s'=0$, it is
\be
[X^k(\t,0),X^l(\t,0)]=2\pi i\a'(M^{-1}\cF)^{kl},
\ee
and that when $\s=\s'=\pi$, it is
\be
[X^k(\t,\pi),X^l(\t,\pi)]=-2\pi i\a'(M^{-1}\cF)^{kl}.
\ee

The infinite series on the right hand side of (\ref{XP})
\be
\frac{1}{\pi} (1+ \sum_{n\neq 0}\cos n\s \cos n\s')
\ee
is the Fourier expansion of the delta function
$\delta(\s-\s')$ for functions defined on the interval
$[0,\pi]$ with vanishing derivatives on the boundary.

Thus we see that the commutation relations are the
standard ones for any point in the interior of the open string.
At the two end points of the open string where the D-branes sit,
we find the spacetime coordinates to be noncommutative.

The noncommutativity of the spacetime coordinates
of an open string was first suggested in ref.\cite{AAS}.
However the commutation relation they found was
\be \label{AASJ}
[X^i(\t,\s),X^j(\t,\s')]=2\pi i\cF^{ij}\th(\s-\s').
\ee
This expression is not well defined for $\s=\s'$ and it
does not agree with our result.

Note that $(M^{-1}\cF)^{kl}$ is anti-symmetric as required.
These are different from the usual commutation relations
for free fields. Since the end-point of the string can be
identified with the D-brane worldvolume, we will use this
relation in the next section for the points $\s=0,\pi$
to make statements about the noncommutativity on the
D-brane worldvolume and to compare them with what we know
from M(atrix) model compactification.

In general, one can define for a D$p$-brane the following number operator,
\be
N_n=N'_n+ N''_n, \quad n \neq 0, 
\ee
where $N'_n= M_{ij} a^i_{-n}a^j_n /(2 \a' n) $
and $N''_n= a^a_{-n}a^a_n /(2 \a' n)$ are the number
operators for the (properly normalized) oscillation modes and the mass $M$
is given by
\be 
M^2 = - P_{tot}^k P_{tot}^k.
\ee

The action (\ref{action}) we started with is obtained from
a diffeomorphic invariant action with background fields
by gauge fixing \cite{DMR,Alam}.
Thus we need to impose the constraints
\be
(\del_{\t}X^{\mu})^2+(\del_{\s}X^{\mu})^2=0, \quad
\del_{\t}X^{\mu}\del_{\s}X_{\mu}=0
\ee
on the physical states.
One can extract from this the Virasoro generators as usual, they are
\be
L_k=\frac{1}{4\a'} 
\sum_{n\in\Z}\left(M_{ij}a^i_{k-n}a^j_{n}+a^a_{k-n}a^a_n\right),
\ee
and they satisfy 
\be
L_m^\dagger = L_{-m} 
\ee 
as a result of \eq{inv}.
With the usual normal ordering of negative modes 
preceding the positive modes,  
the normal-ordered Virasoro generators are defined as 
\be
L_k=\frac{1}{4\a'} 
: \sum_{n\in\Z}\left(M_{ij}a^i_{k-n}a^j_{n}+a^a_{k-n}a^a_n\right) : 
\, .
\ee
It is easy to check that they satisfy the standard Virasoro algebra
\be
[L_m, L_n] = (m-n) L_{m+n} +\frac{d}{12} m(m^2-1) \d_{m+n}, \quad 
d =\, \mbox{spacetime dimension},
\ee
with a central charge not modified by the presence of $\cF$.
This is necessary for our quantization to be consistent.
As usual, one requires that $L_k$ for $k\geq0$ annihilate
a physical state upon quantization.

\section{M(atrix) Model on Torus}

Consider the case of D$2$-branes. {}From (\ref{XX}), 
for $\cF_{01}=\cF_{02}=0$, the open string end-point at $\s=0$ satisfies
\be \label{end0}
[X^1(0),X^2(0)]= 2\pi i \a'\frac{\cF}{1+\cF^2},
\ee
where $\cF=\cF_{12}$.
In the static gauge, the coordinates of the open string
end-point is the same as the coordinates of the D-brane
worldvolume.
Therefore, the noncommutativity (\ref{end0}) for the string
implies that the D-brane worldvolume is a noncommutative space.
We will show now that (\ref{end0}) in fact agrees with the matrix model
results \cite{CDS,DH}.

According to the T-duality in string theory,
a two-torus of radii $R_1,R_2$ and background $B$ field flux of $\th$
is dual to the two-torus of radii $\Sigma_1,\Sigma_2$ with
the background flux of $\tilde{\th}$, where
\beqa
\Sigma_1&=&\frac{l_s^2 R_2}{\sqrt{l_s^4\th^2+R_1^2 R_2^2}},
\label{Sigma1}\\
\Sigma_2&=&\frac{l_s^2 R_1}{\sqrt{l_s^4\th^2+R_1^2 R_2^2}},
\label{Sigma2}\\
\tilde{\th}&=&\frac{-l_s^4\th}{l_s^4\th^2+R_1^2 R_2^2}.
\label{tildeB}
\eeqa
One can derive from this  the relation between the $B$ fields
in the two dual theories
\be \label{BB}
\tilde{B}=\frac{l_s^2\tilde{\th}}{\Sigma_1 \Sigma_2}
=-\frac{l_s^2\th}{R_1 R_2}=-B.
\ee

Consider D0-branes in the background of $\tilde{B}$
on a $T^2$ with radii $\Sigma_i$ as in (\ref{Sigma1}),(\ref{Sigma2}).
By T-duality we get D2-branes on $T^2$ with radii $R_1,R_2$
and a background $B$ field.
The mass of the D0-brane is $(g'_s l_s)^{-1}$,
where $g'_s$ is the string coupling for the dual theory and
$l_s$ is the string length scale.
The mass of the D2-brane according to the DBI action is
\be
M=T_2 V\sqrt{1+B^2},
\ee
where $T_2=\left( (2\pi)^2 g_s l_s^3\right)^{-1}$ is the D2-brane
tension, and $V=(2\pi)^2 R_1 R_2$ is the D2-brane volume.
The duality between the D0-brane and the D2-brane implies that
\be
g'_s=\frac{g_s l_s^2}{R_1 R_2\sqrt{1+B^2}}.
\label{gg'}
\ee

According to the M(atrix) model,
the D2-brane physics is a gauge theory living on a noncommutative torus
\cite{CDS,DH}
\be
[x_1,x_2]=-2\pi i\tilde{\theta},
\ee
where $x_i\in [0,2\pi)$.

On the other hand, in the static gauge we have $X^i=R_i x_i$
as the coordinates on the D$2$-brane, so
\be
[X_1,X_2]=2\pi i l_s^2\frac{B}{1+B^2} 
\ee
on using \eq{Sigma1}, \eq{Sigma2} and \eq{BB}.
This agrees precisely with (\ref{end0}) for $F=0$.

Although in the matrix model derivation we have assumed
the compactification on $T^2$, after we get the
final result we can take the limit of $\Sigma_i\rightarrow 0$
with $B$ fixed.
This corresponds to uncompactified D2-brane worldvolume.
There is an issue about the decompactification.
Since the $B$ field is a gauge field, in an infinite
space we can make a gauge transformation so that $B=0$.
So this seems to say that there is no noncommutativity
in the decompactification limit.
However this conclusion is not completely correct
because on D2-brane the term $\cF=B-F$ is
gauge invariant. So even if $B=0$, there is still
noncommutativity if $\cF\neq 0$.

It may also appear at first sight that there is a mismatch between the
matrix model results and the open string calculation presented here. 
For matrix model compactified on a torus of radii
$\Sigma_1,\Sigma_2$ with a background
flux of $\tilde{\theta}$, the two-dimensional SYM theory lives on a
noncommutative torus of radii $l_s^2/\Sigma_1,l_s^2/\Sigma_2$ 
\cite{CDS,Ho}, which is
different from what is expected from the T-duality of string theory,
i.e. $R_1,R_2$. However this is not a true discrepancy
because the torus on which the SYM theory lives is not exactly
the torus in the dual string theory.
In fact it was shown \cite{Li} that if we consider the
DBI action for D$2$-branes on a torus of radii $R_1,R_2$, its
expansion agrees with a SYM action on a torus with radii
$l_s^2/\Sigma_i$.

The DBI action for a D2-brane in the background $B$ field is
\be
S=T_2 \int dtd^2\xi \sqrt{-\det(G+\cF)},
\ee
where $G_{ij}=\eta_{ij}+\del_i X^a\del_j X_a$ is the induced metric,
and $\xi_i \in [0,2\pi R_i)$ are the coordinates on the D2-brane
worldvolume.
One can expand the Lagrangian with respect to $(1+\cF^2)$ as
\bea
\sqrt{-\det(G+\cF  )}&=&\sqrt{(1+\cF^2)
\left(1-\frac{(\del_i X^a)^2}{1+\cF^2}+\cdots\right)} \nn\\
&=&\sqrt{1+\cF^2}-\frac{1}{2\sqrt{1+\cF^2}}(\del_i X^a)^2+\cdots,
\eea
thus the DBI action contains a part which is of the same form
as the supersymmetric Yang-Mills action
\be
\frac{T_2}{\sqrt{1+\cF^2}}\int dtd^2\xi
\left(\frac{1}{2}(\del_i X^a)^2+\cdots\right).
\label{S1}
\ee

If we start with the dual theory of D0-branes on the dual torus $T^2$
in the context of M(atrix) compactification, we proceed as follows.
The D0-brane SYM action is
\be
S=\frac{1}{g'_s l_s}\int dt
\left(\frac{1}{2}(\del_t X^{\mu})^2+\cdots\right).
\ee
The quotient conditions
\be
U_i^{\dag}X_j U_i=X_j+2\pi\Sigma_j\d_{ij}, \quad i=1,2,
\ee
are solved by
\be
X^i=-i 2\pi\frac{\del}{\del y_i}+A^i(y),
\ee
where $y\in[0,2\pi l_s^2 /\Sigma_i)$.
To compare this result with the above,
set $y_i=\sqrt{1+\cF^2}\xi_i$,
so the SYM action of D0-brane becomes
\be
S=\frac{1}{g'_s l_s}\int dt
\int\frac{d^2\xi}{(2\pi l_s^2 /\Sigma_1)(2\pi l_s^2 /\Sigma_2)}
\left(\frac{1}{2}(\del_i X^a)^2+\cdots\right),
\label{S2}
\ee
where $\del_i=\del/\del\xi_i$.
Using (\ref{gg'}), one can see that (\ref{S1}) agrees with (\ref{S2})
when $F=0$.

It is straightforward to generalize the discussion to D$p$-branes.
We get from (\ref{XX})
\be \label{XX1}
[X^k,X^l]=\pm 2\pi i\a'(M^{-1}\cF)^{kl}
\ee
with the plus (minus) sign corresponding to the end-point at
$\s=0 (\pi)$. This formula tells explicitly how the string
theory data appear in the noncommutativity of the D$p$-brane
worldvolume.
As in the D2-brane case, one can show 
that it agrees with the results of M(atrix) theory
if one uses the T-duality radii
instead of using the M(atrix) model radii natively.  
Let us explain this in more detail. 
Consider a D$p$-brane on $T^p$ with radii $R_i$ in a background
$B$-field. The dimensionless metric and flux are
\be \label{metric1}
G_{ij} = \d_{ij} R_i^2 / 2\a', \quad \theta_{ij} = B_{ij} R_i R_j  /
2\a',
\ee
where $i,j= 1, \cdots p$ and there is no sum of indices. This is the
theory of interest.
The T-dual theory is D0-branes on a $T^p$ with radii $\S_i$
in a background of $\tilde{B}$.
The dimensionless metric $\tilde{G}$  and flux $\tilde{\theta}$ are
(set $2\a'=1$ for convenience)
\be \label{metric2}
\tilde{G} + \tilde{\theta} = (G+\theta)^{-1}.
\ee
See \cite{GPR} for a review of T-duality.
Denote
\be
(G+\theta)_{ij} = R_i R_j (\d_{ij} + B_{ij}) = R_i R_j g_{ij}
\ee
and so
\be
\tilde{G}_{ij} + \tilde{\theta}_{ij} = \frac{1}{R_i R_j} g^{-1}_{ij}.
\ee
It is
\be
g^{-1} = (1+B)^{-1} = (1+B^2 +B^4 + \cdots) -(B+B^3+ \cdots),
\ee
where the first sum is symmetric and the second sum is antisymmetric.
Define
\be
m_i{}^j = \d_i{}^j - B_i{}^k B_k{}^j.
\ee
This is $M$ for the case of $A=0$. Then
\be
g^{-1} = m^{-1} - B m^{-1}.
\ee
Hence
\be
\tilde{G}_{ij}= \frac{1}{R_i R_j}  m^{-1}_{ij}, \label{e1} \quad
\tilde{\theta}_{ij} = - \frac{1}{R_i R_j} (B m^{-1})_{ij}. \label{e2}
\ee

Generalizing the arguments in \cite{CDS,DH}, it is easy to see that 
M(atrix) model predicts in general
\be \label{matrix}
[x^i, x^j] = -2 \pi i \tilde{\theta}^{ij},
\ee
where  $x^i \in(0, 2\pi)$ is the angular coordinates of the
D$p$-brane. Substituting \eq{e2} into \eq{matrix}, we get
\be \label{ours}
[X^i, X^j] = 2\pi i \a' (m^{-1} B)^{ij} \quad\mbox{with}\quad 
X^i = R_i x^i,
\ee
which is precisely our result for the case of $A=0$. 
{}From the D$p$-brane point of view, it is natural to expect that it is
$\cF$ (instead of $B$) that controls the noncommutativity of the 
worldvolume field theory. 
We will now explain the reason for the signs in \eq{XX1}
from the point of view of D-brane worldvolume theory.

\section{D-brane Field Theory}

Consider the D2-brane case for simplicity, 
the other end-point of the open string at $\s=\pi$
satisfies
\be \label{end1}
[X^1(\pi),X^2(\pi)]= - 2 \pi i \a'\frac{\cF}{1+\cF^2}.
\ee
Note that there is only a difference in sign on the right
hand side when compared with (\ref{end0}).
We will now show  that this sign difference is important
for the gauge field theory on the D-brane worldvolume to exist.

Let us summarize (\ref{end0}), (\ref{end1}) as
\be
[x^{\pm}_1, x^{\pm}_2]=ih^{\pm},
\ee
where $x^-_i=X_i(0)$ and $x^+_i=X_i(\pi)$ denote the two end-points
of the open string at $\sigma=0$ and $\sigma=\pi$,
and $h^{\pm}=\pm{2\pi\a'\cF}/(1+\cF^2)$.
The coordinates on the D2-brane can be identified with
the coordinates of the end-points of the open string.
It may seem at first that for both end-points of a string
to end on the same D2-brane we need $h^+ = h^-$ in order
to have a unique commutation relation for the D2
worldvolume coordinates, but below we argue that this
condition should be instead
\be \label{hh}
h^+ = - h^-,
\ee
otherwise we don't know how to describe the D2-brane
gauge field theory on a single noncommutative space.

The end-points of open strings are described by
a complex field on the D2-brane because
there are both positive and negative charges.
For instance the fermionic ground state on the open string is
described in the D2-brane field theory by a fermionic field:
\be
\psi(x)=\sum_n\left(a_n^{\dag}(x)e^{iw_n t}+b_n(x)e^{-iw_n t}\right),
\ee
and similarly for bosonic modes.
The interaction between $\psi$ and $A_i$ on D2-brane is given by
a term like $L_{int}=\bar{\psi} \gamma^i A_i\psi$
in the Lagrangian.
In this term the field $A(x)$ acts on $a^{\dag}(x)$ from
the left and it acts on on $b^{\dag}(x)$ from the right.
To bring the action of the $A_i$ field on the two charges to
the same form, we define $A'$ from $A$ by
\be
A'(x')=\sum A_{mn} e^{inx'_2}e^{imx'_1},
\ee
if $A$ is given by $A(x)=\sum A_{mn}e^{imx_1}e^{inx_2}$.
The coordinates $x'_i$ are operators just like $x_i$.
They are defined by multiplication from the right
\be \label{x'}
x'_i \psi(x)=\psi(x)x_i,
\ee
so $x'$ satisfies the opposite algebra of $x$.
Now the interaction term can be written in terms of
$A$ acting on $a^{\dag}$ and $A'$ acting on $b^{\dag}$
both from the left. In this convention we see that
the gauge field seen by the positively charged end-points,
which correspond to $a^{\dag}$, is $A(x)$; and the
gauge field seen by the negatively charged end-points,
which correspond to $b^{\dag}$, is $A'(x')$.
This indicates that we have $x\sim x^+$ and $x'\sim x^-$.
Since $x'$ is the opposite algebra of $x$, we need (\ref{hh}).

It follows from \eq{x'} that $x'$ commutes with $x$.
This is consistent with the fact that the two end-points
of an open string commute
\be
[X^i(0),X^j(\pi)]=0,
\ee
which follows directly from \eq{XX}.

The physical reason for (\ref{hh}) is just that the two end-points have
opposite charges under $\cF$ and $h^{\pm}$ is proportional to $\cF$.
So the symmetry of charge conjugation is preserved
only if (\ref{hh}) holds.

In the absence of background fields $B$ and $F$, the low energy
physics of $N$ coincident D$p$-branes is given by the $U(N)$
SYM theory dimensionally reduced to $p+1$
dimensions. In ref.\cite{HV2}, it was proposed that
the gauge field theory
for D-branes is also given by SYM theory on a quantum plane when the
background $B$ field exists, just like the case in matrix theory.
Since the matrix model is related to the D-brane physics by the
Seiberg limit \cite{Seib}, it is not at all obvious that this
statement is correct.  In fact, there is a serious mismatch for this
interpretation \cite{HV2}, that is, the $SL(2,\Z)$ transformation for
the $B$ field in such a theory is correct only in the matrix model
limit.  Nevertheless, our analysis of the open string quantization
shows that the proposal of \cite{HV2} must be correct in the sense
that the field theory of D-brane must be a field theory on a
noncommutative space, although we might be still missing something to
fill the noted gap.

\section{Remarks}

\noindent{\bf Fermionic Modes:}

In the above we have only focused on the bosonic modes on the
open string. Now we consider the fermionic modes.
In the RNS string action, the fermionic part is
\be
S_F=-\frac{i}{4\pi\a'}\int d^2\s
(\bar{\psi}^{\mu}\r^{\a}\del_{\a}\psi_{\mu}).
\ee
It is not modified by a constant $B$ field.
The supersymmetry on the world sheet is
\be
\d X^{\mu}=\bar{\epsilon}\psi^{\mu}, \quad
\d\psi^{\mu}=-i\r^{\a}\del_{\a}X^{\mu}\epsilon.
\ee
The equations of motion
\be
\r^{\a}\del_{\a}\psi^{\mu}=0,
\ee
and the boundary conditions
\bea
&\psi^{\mu}_+=\pm\psi^{\mu}_-
\eea
at $\s=0,\pi$ are the same as before.
Therefore the fermionic fields on the open string are
exactly the same as when $\cF$ is zero.
The two choices of sign in the boundary conditions
lead to the Ramond sector and the Neveu-Schwarz sector
which give the ground states of a spinor and a vector,
respectively. They correspond to the fermionic field
and the gauge field in the SYM theory in $9+1$ dimensions.
For the D$p$-brane field theory the gauge field is
dimensionally reduced to the gauge field $A_i$ and the Higgs $X_a$.
It is therefore quite trivial to include the fermionic modes in all our
discussion above.

\noindent{\bf Generalizations:}

It is clear from our derivation
that the noncommutativity can exist for
generic geometry or topology of the D$p$ brane. 
For example, a D2-brane can be a noncommutative $S^2$ or $S^{1,1}$.
For these cases the commutation relations for the oscillation modes
may be difficult to derive, but for the lowest modes we may use symmetry
requirements to fix their relations, assuming that the $\cF$ field
also respects part of the isometry group. For instance,
for a spherical D$2$-brane, the end-point of an open string 
in Cartesian coordinates satisfies
\be
[X_i,X_j]=ih\eps_{ijk}X_k,
\ee
where $h = h(\cF)$ is a constant satisfying $h(0)=0$.
The point is that the $\cF$ field determines the symplectic
structure and if it respects the isometry of the space
it may be fixed by the symmetry up to an overall factor.

In this paper, we considered string ending on a D$p$-brane and showed
that the end-points become noncommutative in background fields.
Using S-duality, one can turn this into a configuration of a D1-brane
ending on a D$p$-brane with a background R-R $B$-field. 
The end-points of the D1-brane and hence the worldvolume of the
D$p$-brane will again have noncommutative coordinates.
Combining S-duality and T-duality, one can  arrive at more general 
configurations of branes ending on branes. For example, 
a solitonic brane ending on a D-brane; or a D$p$-brane ending
on a D$q$-brane. 
In these cases, the end-points of the ``smaller'' brane and hence the
``host brane'' worldvolume will again have noncommutative coordinates,
although showing it directly may be difficult due to complications in
quantization of higher dimensional branes and solving the
corresponding boundary conditions which are no longer linear for
higher dimensional extended objects.
Nevertheless, the lowest energy modes can still be dealt with.
Let us discuss some features in a similar classical 
spirit as in \cite{Duff}.

For instance for a membrane ending on a M5-brane, the
boundary conditions analogous to (\ref{BC1}) are
\be
\del_2 X^i-\cF^i{}_{jk}\del_0 X^j\del_1 X^k =0 ,
\ee
where $a,b =0,1,2$ are the membrane worldvolume index.
This boundary condition can be derived from the bosonic action
\be
S=\int d^3\s \left(\frac{1}{2}(\del_a X^i)^2-
\frac{1}{6}\eps^{abc}\cF_{ijk}\del_a X^i\del_b X^j\del_c X^k\right).
\ee
At the boundary of the membrane, $X^k$ can be identified with
the M5-brane worldvolume and
$\cF$ is the modified field strength on the M5-brane.
The nonlinearity of the boundary conditions makes it difficult
to obtain a generic solution of $X^i$.
It is however possible to examine the lowest energy modes
\be
X^i=x^i+p^i_a\s_a+\cF^i{}_{jk}p^j_0 p^k_1\s_2 + \cdots .
\ee
Choosing the range of $\s_2$ to be $[-\pi/2, \pi/2]$, one finds
the $\Omt$ for these modes
\be
\tilde{\Omega}=\left(\bfd p_0^i-
\cF^i{}_{jk}\cF^k_{lm}\bfd(p_0^l p_1^j p_1^m)\right)\bfd x_i.
\ee
One can check that when $X^1$ is compactified on a circle
of radius $R$, this reduces to the case
for an open string with $\cF_{ij}=\cF_{ijk}p^k_1$.

\noindent{\bf Uncertainty Relation:}

In the second quantization of string theory, one has to 
integrate over all possible configurations.
So even if we consider a classical background of vanishing VEV for $\cF$, 
in general 
\be \label{stringfield}
\int[D\cF][D\Psi]e^{-S(\cF,\Psi)}(\Delta X_1)^2(\Delta X_2)^2>0,
\ee
will not be zero and noncommutativity on the worldvolume of extended objects 
is a generic feature of string theory. 
In \eq{stringfield},  $S$ is the action of string field theory.  
$\Psi$ is the wave function for D-brane and 
$(\Delta X_i)^2$ and $\la X_i(\sigma,t)\ra$ are 
defined by
\be
(\Delta X_i)^2=\int[DX(\sigma,t)]\Psi(X)^{\dag}
(X_i(\sigma,t)-\la X_i(\sigma,t)\ra)^2\Psi(X),
\ee 
\be
\la X_i(\sigma,t)\ra=\int[DX(\sigma,t)]\Psi^{\dag}X_i(\sigma,t)\Psi.
\ee
It would be interesting \cite{work} if some connection of this with
the work of \cite{Yon} can be established. 

\noindent{\bf Conclusion:}

In this paper, we show that noncommutativity of the D-brane
worldvolume is a direct consequence of quantizing open string theory
in non-zero backgrounds of $B - F$. 
The quantization we proposed is a consistent quantization which
covers and generalizes the usual situations, and the unitary
flow equation is reproduced. Moreover, we show that
it agrees with the noncommutativity 
obtained in M(atrix) model compactification
provided we use the T-duality radii.
Our results however generalize the M(atrix) model results
in several ways.

In M(atrix) model, noncommutativity was argued to exist on
compactification. Here, we showed that not only that we
don't need compactification for noncommutativity,
in general we have noncommutativity whenever $B-F$
is nonzero on a D-brane. 
Also we give explicit result (\ref{XX1}) of how
noncommutativity should look like for higher dimensional D$p$-brane. 
The data invloved is the gauge invariant $\cF$ instead of the $B$ as
suggested in M(atrix) model compactification.
It would be very interesting to see this directly from the M(atrix)
model compactification or other point of view \cite{work}.

In a string field theory, one has a Hilbert space that
contains all the states of string theory, including the closed strings,
open strings and D-branes. A closed string can break itself and end
on a D-brane. It can also break itself and interact with another
closed string or open string.
One may perhaps anticipate that noncommutativity
is not confined to the worldvolume of D-branes, but is generic to all 
extended objects in a consistent formulation of string theory.

\section*{Acknowledgment}

C.S.C. thanks A. Bilal, M. Cederwall, C. Hull
and R. Russo for helpful discussions.
P.M.H. thanks M. M. Sheikh-Jabbari for valuable discussions
and the Asia Pacific Center for Theoretical Physics
for hospitality where part of this work was done.
The work of C.S.C. is supported by the Swiss National
Science Foundation. The work of P.M.H. is supported in part by
the National Science Council, Taiwan, R.O.C.

\ed